\documentclass[paper]{JHEP}
\usepackage{epsfig}
\usepackage{psfrag}
\newcommand{\be}{\begin{equation}}
\newcommand{\ee}{\end{equation}}
\newcommand{\bea}{\begin{eqnarray}}
\newcommand{\eea}{\end{eqnarray}}

\title{\boldmath Radiative $\phi$$-$meson decays and $\eta$-$\eta^\prime$
mixing:
a QCD sum rule analysis \unboldmath}

\author{Fulvia De Fazio\\
Centre for Particle Theory, University of Durham\\
Durham DH1 3LE, U.K.\\
E-mail: \email{Fulvia.de-Fazio@durham.ac.uk}}

\author{M.R. Pennington\\
Centre for Particle Theory, University of Durham\\
Durham DH1 3LE, U.K.\\
E-mail: \email{M.R.Pennington@durham.ac.uk}}

\abstract{The radiative transitions
$\phi \to \eta \gamma$ and $\phi \to \eta^\prime \gamma$ 
are analysed using QCD sum-rules. 
  At  leading order in perturbative QCD, we obtain the results:  
${\cal B}( \phi  \to \eta \gamma)=(1.15 \pm 0.2)~10^{-2}$ and  
${\cal B}( \phi \to \eta^\prime \gamma)=(1.18 \pm 0.4)~
10^{-4}$,  in very good agreement
with existing experimental data.
The related issue of  $\eta-\eta^\prime$ mixing is discussed and we give
predictions for the $\eta$ and $\eta^\prime$ decay constants in the framework
of a  mixing scheme in the quark-flavour basis. }

\keywords{Electromagnetic processes and
properties, Sum Rules, Chiral Lagrangians}

\preprint{DPT/00/14\\
\hepph{0006007}}

\begin{document}

\section{Introduction}

Radiative $\phi$ meson decays represent an important source of information 
on low-energy hadron physics, shedding light, for example, 
on the structure and properties of low-mass resonances, such as the $f_0(980)$.
In particular, radiative $\phi$ decays to $\eta$ and
$\eta^\prime$ can provide insights into the long standing problem of
$\eta-\eta^\prime$ mixing and probe the strange quark content of
the light pseudoscalars \cite{eilam,rosner}.
Radiative
$\phi$ decays, not only raise interesting theoretical issues,
but are an important focus of the data-taking by the KLOE experiment 
at the DA$\Phi$NE
$\phi$-factory  \cite{book}, where a
large sample of $\phi$ decays will be collected, dramatically improving
the experimental information already obtained by the VEPP$-$2M groups at
Novosibirsk 
\cite{novos}.

This paper is devoted to
the analysis of the radiative $\phi \to \eta \gamma$ and 
$\phi \to \eta^\prime \gamma$ transitions.
In contrast to the light vector meson case, where the  $\omega$ and
$\phi$ are recognised as almost ideally mixed states  with
quark content of well defined flavour, $\eta-\eta^\prime$ mixing is
still a much debated subject. The once conventional description was to
adopt a single mixing
angle in the octet-singlet flavour basis. Various attempts to estimate such
an angle lead to results ranging from $-10^0$ to $-20^0$
\cite{feldrev}. More recently, Leutwyler et al. \cite{leutproc,kaiser} have
shown
that a consistent treatment of the $\eta-\eta^\prime$ system 
requires the introduction of two mixing angles with a consequent
redefinition of the particle decay constants. An equivalent description,
as explained in more detail below, 
is obtained if the mixing basis is chosen to be the
quark-flavour basis instead of the octet-singlet one \cite{feld}. In such a
scheme, it has been shown that a description in terms of a single 
mixing angle is quite reliable leading to predictions satisfying  
constraints from
Chiral Perturbation Theory \cite{leutproc}. 

In this context a central role is played by the $U(1)_A$ anomaly. 
Since the flavour-singlet axial vector
current is not conserved due to this anomaly, the $\eta^\prime$ meson
cannot be identified as the ninth Goldstone boson. This 
crudely explains the
fact that the $\eta^\prime$ is much heavier 
than the other members of the pseudoscalar nonet. 
By combining chiral symmetry with the concepts of the large 
$N_c$ limit of QCD, Leutwyler\cite{leut} has extended 
Chiral Perturbation Theory from an expansion in the light quark
masses and momenta to encompass powers of
${1/ N_c}$. Then in the limit $N_c \to 
\infty$, the $U(1)_A$ anomaly vanishes and the $\eta^\prime$ can
formally be identified with the ninth Goldstone boson. In order to make
the framework more predictive and closer to reality, correction terms are
added to the effective  Lagrangian of the theory, expressing the deviation
from the chiral limit for the decay constants and the masses of the light
mesons. 
A Wess-Zumino-Witten (WZW) term describing the anomalous coupling to
photons
 must also be added. Radiative $\phi$
decays provide additional information on the strength of this WZW term.

In the following we  analyse $\phi$ radiative decays using QCD
sum-rules \cite{shifman} at leading order in perturbative QCD. These
sum-rules are widely recognised 
as a reliable technique for including the effects of non-perturbative QCD. 
In section 2 we  survey possible $\eta-\eta'$ mixing schemes, with
particular emphasis on  the flavour basis mixing scheme developed by
Feldmann et al. \cite{feld}. We  give the relation between the
parameters
characterising such a scheme and those in the octet-singlet basis. In
section 3, as a
preliminary to our analysis of $\phi$ radiative decays, we 
compute  the coupling of the strange pseudoscalar current
both to the $\eta$ and $\eta^\prime$ mesons using two-point QCD sum-rules. 
These provide  key inputs into the three-point QCD sum-rule
developed to estimate the decay widths
$\Gamma(\phi \to \eta \gamma)$ and 
$\Gamma(\phi \to \eta^\prime \gamma)$, in section 4. The possible effect
of ${\cal O}(\alpha_s)$ corrections is discussed at the end of this
section.

Though we often refer to the problem of mixing  and we actually
work with  interpolating currents defined in the quark flavour basis,
our results do not depend on any specific mixing scheme and so provide
a genuinely mixing scheme independent set of QCD predictions. 
If then a particular
flavour mixing scheme is adopted, our results can be translated into a
prediction for one of the mixing angles.
In section 5 we again use two-point QCD sum-rules to compute the
coupling of $\eta$ and $\eta^\prime$  to the strange and non-strange axial
currents, identifying the results with the decay constants in the flavour
basis mixing scheme. The results again allow us to estimate the
mixing parameters in such a scheme.
The results in sections 3 and 5 can be exploited to obtain an estimate of
the contribution of the anomaly to the couplings computed in section 3.
In section 6 we draw our conclusions.

\section{On  $\eta-\eta^\prime$ Mixing}
Let us first recall  the usual parametrization of 
$\eta-\eta^\prime$ mixing in the octet-singlet basis.  
We define current-particle matrix elements as
\be
<0|J_{5 \mu }^i|P(p)>=i ~f_P^i p_\mu \;\;\; (i=8,0; ~~~
P=\eta,\eta^\prime) \;, 
\label{lepconst} \ee
\noindent with $J_{5 \mu }^8$ the $SU(3)_F$ octet axial vector
current:
\be
J_{5 \mu }^8={1 \over \sqrt{6}} \left( {\overline u} \gamma_\mu \gamma_5 u +
{\overline d} \gamma_\mu \gamma_5 d -2  {\overline s} \gamma_\mu \gamma_5 s 
\right)
\label{octvec} \ee
\noindent and $J_{5 \mu }^0$  the singlet current:
\be
J_{5 \mu }^0={1 \over \sqrt{3}} \left( {\overline u} \gamma_\mu \gamma_5 u +
 {\overline d} \gamma_\mu \gamma_5 d +  {\overline s} \gamma_\mu \gamma_5 s
\right)\; .
\label{singvec}\ee
\noindent 
As already mentioned, two
mixing angles, $\theta_8$ and $\theta_0$,
are required \cite{leut} in order to treat 
mixing consistently.  Accordingly, the couplings in (\ref{lepconst})
can be defined as follows:
\bea
f_\eta^8=f_8 ~\cos \theta_8 && \hskip 1 cm f_\eta^0=-f_0 ~ \sin \theta_0
\nonumber \\
f_{\eta^\prime}^8=f_8 ~\sin \theta_8 && \hskip 1 cm f_{\eta^\prime}^0=f_0 ~
\cos \theta_0\; .
\label{mixoctsin}
\eea
\noindent
Alternatively we can consider two independent
axial vector currents with distinct quark flavour:
\bea
J_{5 \mu }^q &=& {1 \over \sqrt{2}}({\overline u} \gamma_\mu \gamma_5 u+
{\overline d} \gamma_\mu \gamma_5 d) \nonumber \\
J_{5 \mu }^s&=& {\overline s} \gamma_\mu \gamma_5 s \; .
\label{currents}
\eea
\noindent
The couplings of the $\eta$ and $\eta^\prime$ mesons to the currents 
(\ref{currents}) can
be defined analogously to (\ref{lepconst}). The decay constants are
written  according to  the following mixing pattern:
\bea
f_\eta^q=f_q ~\cos \phi_q && \hskip 1 cm f_\eta^s=-f_s ~ \sin \phi_s
\nonumber \\
f_{\eta^\prime}^q=f_q ~\sin \phi_q && \hskip 1 cm f_{\eta^\prime}^s=f_s ~
\cos \phi_s \;\;\;\;.
\label{constmix}
\eea

Though there are, of course, two angles in each basis, 
Feldmann \cite{feld} has shown that the mixing is   specified
quite accurately in terms of a
single mixing angle, i.e. $\phi_q=\phi_s=\phi$, since
$|\phi_s-\phi_q|/ (\phi_s+\phi_q)\ll 1$,
resulting in a much simpler
framework. In this approximation, 
the states follow the same mixing pattern as the decay constants:
\bea
|\eta> &=& \cos \phi ~ |\eta_q> -\sin \phi ~|\eta_s> \nonumber \\
|\eta^\prime> &=& \sin \phi ~ |\eta_q> +\cos \phi~ |\eta_s> \label{statemix}
\eea
where $|\eta_q>$ and $|\eta_s>$ have a quark content defined by ideal 
mixing.
We will refer to this simply as mixing in the quark-flavour basis, in
order to distinguish it from the previous one, which will be referred to
as  mixing in the octet-singlet basis. 
It is straightforward to obtain the relations  between the
parameters in the two mixing schemes: 
\be
\tan~\theta_8={f_q\, \sin \phi_q -\sqrt{2} f_s\, \cos \phi_s \over 
f_q\, \cos \phi_q +\sqrt{2} f_s \,\sin \phi_s} 
\;\;\; , \;\;\;\; 
\tan~\theta_0={f_s\,\sin \phi_s -\sqrt{2} f_q \, \cos \phi_q \over
f_s\, \cos \phi_s +\sqrt{2} f_q \,\sin \phi_q}
\label{twoangles}
\ee
which become, for $\phi_q=\phi_s=\phi$  \cite{feld}:
\be
 \theta_8=\phi-\arctan \left( {\sqrt{2} f_s \over f_q} \right) \hskip 1 cm 
\theta_0=\phi-\arctan \left( {\sqrt{2} f_q \over f_s} \right) 
\; . \label{angles}
\ee
Moreover:
\bea
f_8^{\,2} &=& {1 \over 3} f_q^{\,2} +{2 \over 3} f_s^{\,2}+
{2 \sqrt{2} \over 3} f_q f_s
\sin(\phi_s-\phi_q) \nonumber \\
f_0^{\,2} &=& {2 \over 3} f_q^{\,2} +{1 \over 3} f_s^{\,2}-
{2 \sqrt{2} \over 3} f_q f_s
\sin(\phi_s-\phi_q) \;\;.
\label{f08}
\eea
In the following sections we derive  the ingredients
necessary for the description of the radiative $\phi \to \eta \gamma$ and
$\phi \to \eta^\prime \gamma$ decays, without 
assuming any specific mixing framework. 
Then, in
section 5, we will evaluate the couplings of the $\eta$ and $\eta^\prime$
to
the axial vector current and estimate the parameters appearing in 
(\ref{constmix}), obtaining a prediction for $\phi_q$, $\phi_s$. At the end,
we  shall comment on the accuracy of the approximation $\phi_q=\phi_s$.

\section{Two-point function for $\eta$ and 
$\eta^\prime$ couplings to the pseudoscalar current}

Let us consider  the matrix element of the divergence of the axial-vector
current:
\be
<0|\partial^\mu J_{5 \mu }^s|\eta>=m_\eta^{\,2} f_\eta^s\;.
\ee

\noindent As is well known,  this divergence contains the axial-vector
anomaly:
\be
\partial^\mu J_{5 \mu }^s=\partial^\mu ({\overline s} \gamma_\mu \gamma_5 s)=
2 m_s {\overline s} i\gamma_5 s +{\alpha_s \over 4 \pi} G \tilde G \; ,
\label{anomaly}
\ee

\noindent where $G$ is the gluon field strength tensor and ${\tilde G}$ 
its dual. 
This gives a relation between the matrix elements of the axial-vector
current and of the pseudoscalar current:
\be
2 m_s <0|{\overline s} i\gamma_5 s| \eta^{(\prime)}>=f_{\eta^{(\prime)}}^s
m^2_{\eta^{(\prime)}}-\langle 0 \left|{\alpha_s \over 4 \pi}G \tilde
G\right|\eta^{(\prime)} \rangle
\;. \label{glue}
\ee
 
 Let us call:
\be
 <0|{\overline s} i\gamma_5 s|\eta>=A \;\;\;, \label{adef}
\ee
and compute this
quantity by QCD sum-rules starting from the two-point correlator:
\be
T_A(q^2)=i ~\int d^4 x e^{iq \cdot x} <0|T[ J_{5}^s(x) J_{5}^{s
\dagger}(0)]|0>
\label{cor_a}
\ee

\noindent
where $J_5^s={\overline s} i\gamma_5 s$. The correlator (\ref{cor_a})
is given by the dispersive representation:
\be
T_A(q^2)={1 \over \pi} \int_{4 m_s^2}^\infty \, ds\;{\rho(s) \over s-q^2}
+{\rm subtractions} \;. 
\label{disp}
\ee
\noindent In the region of low values of $s$, the physical spectral density
contains a $\delta-$function term corresponding to the coupling of
the $\eta$ to the pseudoscalar current. Picking up this contribution, we
can write (dropping possible subtractions which we discuss later):
\be
T_A(q^2)={ A^2 \over m_\eta^2 -q^2}+ {1 \over \pi} \int_{s_0}^\infty\,ds\;
{\rho^{had}(s) \over s-q^2} \;.
\ee
\noindent This corresponds to assuming that the contribution of higher
resonances and continuum of states starts from an effective threshold
$s_0$.
On the other hand, the correlator $T_A(q^2)$  can be computed in QCD
by expanding the $T$-product in (\ref{cor_a}) by an Operator Product
Expansion (OPE) as the sum of a perturbative contribution plus 
non-perturbative terms which are proportional to vacuum expectation values of
quark and gluon gauge-invariant operators of increasing dimension, the so
called vacuum condensates. In practice, only a few condensates are
included, the most important contributions coming from  the dimension 3
$<{\overline q} q>$ and dimension 5 $<{\overline q}g \sigma  G q>$. Here we 
follow such a prescription.

In the QCD expression for the two-point correlator considered, the
perturbative term can also be written dispersively, so that:
\be
T_A^{QCD}(q^2)={1 \over \pi} \int_{4 m_s^2}^\infty\,ds\; {\rho^{QCD}(s) \over
s-q^2} + d_3 <{\overline s } s> +d_5<{\overline s}g \sigma  G s>+...
\label{qcd_a} \; ,
\ee
\noindent where the spectral function $\rho^{QCD}$ and the coefficients
$d_3$, $d_5$ can be computed in QCD.

The next step consists in assuming quark-hadron duality, which amounts to
assuming the physical and the perturbative spectral density are dual to
each other, in the sense that they should give the same result 
when integrated appropriately above some $s_0$. This leads to the sum-rule:
\be
{ A^2 \over m_\eta^2 -q^2}={1 \over \pi} \int_{4 m_s^2}^{s_0}\,ds\; 
{\rho^{QCD}(s) \over s-q^2} + d_3 <{\overline s } s> +d_5<{\overline s}g \sigma G s>+...
\label{sr} 
\ee
\noindent This expression can be improved by applying 
to both sides of (\ref{sr}) a  Borel transform, defined as follows:
\be
{\cal B} [f(Q^2)]=lim_{Q^2 \to \infty, \; n \to \infty, \; {Q^2 \over n}=M^2}\;
{1 \over (n-1)!} (-Q^2)^n \left({d \over dQ^2} \right)^n f(Q^2) \; ,
\label{tborel}
\ee
\noindent where $f$ is a generic function of $Q^2=-q^2$. The application of
such a procedure to the sum-rules amounts to exploiting the following
result:
\be
{\cal B} \left[ { 1 \over (s+Q^2)^n } \right]={e^{-s/M^2} \over (M^2)^n}
{1 \over (n-1)!} \; , \label{bor}
\ee
\noindent where $M^2$ is known as the Borel parameter. This operation
improves the
convergence of the series in the OPE by factorials and, for suitably chosen
values of $M^2$, enhances the contribution of low lying states. Moreover,
since the Borel transform of a polynomial vanishes, it is correct to
neglect subtraction terms in (\ref{disp}), which are polynomials
in $q^2$.
The final sum-rule reads:
\bea
&& A^2 e^{-{m_\eta^2 \over M^2}}= {3 \over 8 \pi^2} \int_{4 m_s^2}^{s_0}
ds~s ~\sqrt{1-{4 m_s^2 \over s}}e^{-{s \over M^2}}
\nonumber \\
&-& m_s  e^{-{m_s^2 \over M^2}} \Bigg[ <{\overline s} s>\Bigg( 1-{m_s^2 \over
M^2}+
{m_s^4 \over M^4} \Bigg)+ {1 \over M^2} <{\overline s}g \sigma G s> 
\Bigg( 1-{ m_s^2 \over 2 M^2} \Bigg) \Bigg] 
\label{finsr} \; .
\eea

\noindent In the  numerical evaluation of (\ref{finsr})
we use $<{\overline s}s>=0.8 <{\overline q}q>$,  
$<{\overline q}q>=(-0.24)^3$ GeV$^3$, $<{\overline s}g \sigma G
s>=0.8~$GeV$^2 <{\overline s}s>$,
$m_\eta=0.548$ GeV.

\EPSFIGURE{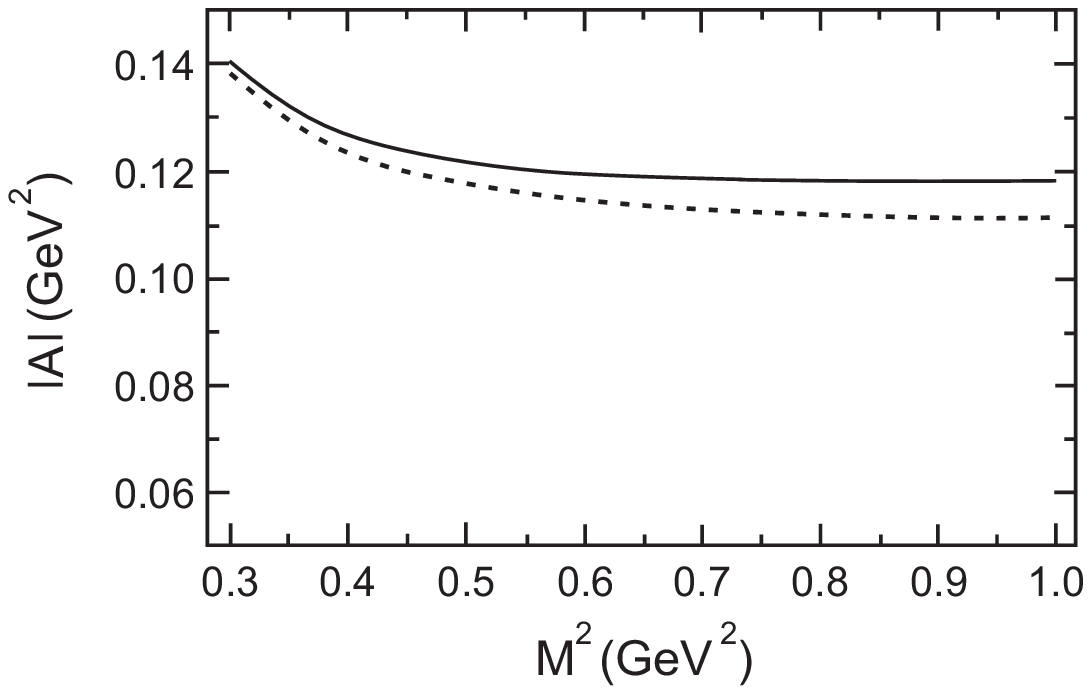,width=7.2cm}
{\label{fig:a}
Coupling of the $\eta$ to the pseudoscalar current as a function of the Borel parameter $M$, for $m_s=0.133$ GeV. The
solid  curve
corresponds to the higher threshold $s_0=(0.95\;{\rm GeV})^2$, the dashed 
curve corresponds to $s_0=(0.9\;{\rm GeV})^2$.}

 The strange quark mass is chosen in the
range $m_s=0.125-0.140$ GeV,
obtained in the same QCD sum-rule framework \cite{noistrano}. 
The threshold is chosen  
below the $\eta^\prime$ pole and varied between 
$s_0=0.9^2-0.95^2$ GeV$^2$. 

Since the Borel parameter has no physical meaning, we require that the
result does not depend on it. This is achieved by finding a \lq\lq stability
window", i.e. an interval of values of $M^2$, where the outcome of the 
sum-rule is almost independent on $M^2$. Such a window is usually sought
in a restricted interval of values of the Borel parameter chosen
 by requiring that the perturbative
contribution is at least 20
$\%$ of the continuum (which corresponds to considering the integral in the
perturbative term up to infinity rather than up to $s_0$), which produces
an upper bound on the Borel parameter: $M^2 \le 1$ GeV$^2$. Additionally
requiring that the perturbative term is greater than the non-perturbative
contribution,  the lower bound: $M^2 \ge 0.5$ GeV$^2$, is obtained.  Then, the
stability window for  $M^2$ in $[0.8,1]$ GeV$^2$ can be selected.

In figure~\ref{fig:a} we plot the sum-rule (\ref{finsr}) for 
$m_s=0.133$ GeV, which  corresponds to the
central value of the range  of values adopted in the analysis.
Taking into account  the uncertainty on $m_s$, we obtain:
\be
|A|=(0.115 \pm 0.004)\; {\rm GeV}^2 \label{a} \; .
\ee
\noindent Some comments are in order on the accuracy of the result
(\ref{a}). This has been obtained at  leading order in 
perturbative QCD, as with all the results presented in
this paper. Consequently, the uncertainty affecting the
determination
(\ref{a}) should be taken {\it modulo} the neglect of 
$\alpha_s$ corrections, the role of which we comment on later.

 Another source of uncertainty is linked to the choice of the strange
quark mass. It should be stressed that the value of this
parameter is quite controversial. On the one hand, lattice determinations
seem to point towards lower values of $m_s$ (for recent reviews see
e.g. \cite{lat}); on the other hand, results obtained in other approaches
 indicate higher values \cite{pich}. As for QCD sum
rules, the result in \cite{noistrano} 
 exploits an  accurate determination of the hadronic
spectral function based on experimental information on the $K \pi$ system
including a non-resonant component in addition to the resonances  in
the
$I=1/2$ channel. In \cite{noistrano} it has been shown that the effect of
this non-resonant contribution is a reduction in the spectral function,
with a consequent
lowering of the value of $m_s$ with respect to  previous QCD sum rule
determinations \cite{oldsr}.  We have therefore chosen to adopt the result 
of
\cite{noistrano} at first for consistency, i.e. using a result obtained by
the same technique, but also because such a value for $m_s$ falls in the
middle of
the existing  range. Moreover, the value ${\bar
m}_s$(1 GeV)=0.125 GeV obtained in \cite{noistrano} could be considered
as a lower
bound on this parameter, since further experimental information could be
added to  improve the sum rule further. Consistently, we have used the
range for ${\bar m}_s$ of 0.125-0.140 GeV quoted
above. The result (\ref{a}) turns
out
to be quite stable. Indeed, we have explicitly checked that 
using still higher values for $m_s$ (up to $0.160$ GeV) would produce
little change.

Let us now consider:
\be
<0|{\overline s} i\gamma_5 s|\eta^\prime>=A^\prime \;\;\;. \label{apdef}
\ee
An analogous calculation gives:
\bea
&& (A^\prime)^2e^{-{m_{\eta^\prime}^2 \over M^2}} + A^2 e^{-{m_\eta^2
\over M^2}}= {3 \over 8 \pi^2} \int_{4 m_s^2}^{s_0^\prime} ds
~s ~\sqrt{1-{4 m_s^2 \over s}}e^{-{s \over M^2}}
\nonumber \\
&-& m_s  e^{-{m_s^2 \over M^2}} \Bigg[
<{\overline s} s>\Bigg( 1-{m_s^2 \over M^2}+
{m_s^4 \over M^4} \Bigg)+ {1 \over M^2}<{\overline s}g \sigma G s> 
\Bigg( 1-{ m_s^2 \over 2 M^2} \Bigg) \Bigg] \;\;\; ,
\eea
\noindent where we have raised the effective threshold up to $s_0^\prime$
in
such a way as to pick up the $\eta^\prime$ pole
too: $s_0^\prime=(1.44,1.55)$ GeV$^2$.
Using $m_{\eta^\prime}=0.958$ GeV and fixing
the stability window for $M^2$ to be $[1.2,2]$ GeV$^2$, we obtain: 

\be
|A^\prime|=(0.151 \pm 0.015)\; {\rm GeV}^2 \; .
\label{ap}
\ee

\noindent We shall use the results (\ref{a}), (\ref{ap}) in the next
section. Though we cannot actually establish the sign of $A$,
$A^\prime$ from the sum rule, we assume that $A \cdot A^\prime >0$.

\section{Radiative $\phi \to \eta \gamma$ and 
 $\phi \to \eta^\prime \gamma$ decays}

Having found the key matrix elements $A$ and $A'$ of
(\ref{a}),   (\ref{ap}),   
we now consider the three-point functions defined by
\be
<\eta(q_2)|{\overline s} \gamma^\nu s|\phi(q_1,\epsilon_1)>=F(q^2)~
\epsilon^{\nu \alpha \beta \delta} (q_1)_\alpha (q_2)_\beta
(\epsilon_1)_\delta
\ee
($q=q_1-q_2$).
In order  to compute the $\phi \to \eta \gamma$  decay, we need the
coupling $g=-{1 \over 3} F(0)$, obtained for a real photon coupling to a
strange quark.
We  consider the three-point function:
\be
\Pi_{\mu \nu}(q_1^2,q_2^2,q^2)=i^2 \int d^4 x~ d^4 y~ e^{-i q_1 \cdot x}\,
e^{i q_2 \cdot y} <0|T[J_5^s(y) J_\nu(0) J_\mu(x)]|0> \label{cor-phi}
\ee
where $J_5^s$ has been defined above and $J_\nu={\overline s} \gamma_\nu s$ is
the vector current. The correlator (\ref{cor-phi}) can be
written as:
\be
\Pi_{\mu \nu}(q_1^2,q_2^2,q^2)=\Pi(q_1^2,q_2^2,q^2)\, 
\epsilon_{\mu \nu \alpha \beta} 
(q_1)^\alpha (q_2)^\beta \label{pi} 
\ee
and a QCD sum-rule can be built up for the structure
$\Pi(q_1^2,q_2^2,q^2)$.
The method closely follows the one described for the two-point sum-rule. 
We assume $\Pi(q_1^2,q_2^2,q^2)$ obeys a dispersion relation
in both
the variables $q_1^2,q_2^2$:
\be
 \Pi(q_1^2,q_2^2,q^2)={1 \over \pi^2} \int ds_1 \int ds_2
{\rho(s_1,s_2,q^2) \over (s_1 -q_1^2)(s_2-q_2^2)} \; ,
\ee
\noindent with possible subtractions.
Such a representation is true at each order in perturbation theory and, as
is 
standard in QCD sum rule analyses, it is assumed to hold in general.
In this case the spectral function
contains,
for low values of $s_1$,
$s_2$, a double $\delta-$function corresponding to the transition $\phi \to
\eta$. Extracting this contribution, we can write:
\be
 \Pi(q_1^2,q_2^2,q^2)={A F(q^2) m_\phi f_\phi \over
(m_\phi^2-q_1^2)(m_\eta^2-q_2^2)} +{1 \over \pi^2}\int_{s_{01}}^\infty d
s_1 \int_{s_{02}}^\infty d s_2\, {\rho^{had}(s_1,s_2,q^2) \over (s_1
-q_1^2)(s_2-q_2^2)} \; ,
\ee
\noindent where subtractions are neglected as later they will vanish on
taking a Borel transform.
The parameter $A$ appearing in the previous equation is just the coupling
of the
$\eta$ to the pseudoscalar current, computed in section~3.
Deriving an OPE-based QCD expansion for $\Pi$ for large and negative
$q_1^2$, $q_2^2$ and $q^2$,  one can write:
\bea
\Pi(q_1^2,q_2^2,q^2)&=&{1 \over \pi^2}\int_{4 m_s^2}^\infty d s_1 \int_{4
m_s^2}^\infty d s_2 
{\rho^{QCD}(s_1,s_2,q^2) \over (s_1 -q_1^2)(s_2-q_2^2)} \nonumber \\
&+& c_3 <{\overline s}
s>+c_5 <{\overline s} g \sigma G s>+...\;\;. 
\eea
Invoking quark-hadron global duality as before, we arrive at the sum-rule:
\bea
{A F(q^2) m_\phi f_\phi \over (m_\phi^2-q_1^2)(m_\eta^2-q_2^2)} &=&
{1 \over \pi^2}\int_D d s_1  d s_2
{\rho^{QCD}(s_1,s_2,q^2) \over (s_1 -q_1^2)(s_2-q_2^2)} \nonumber \\
&+& c_3 <{\overline s }
s>+c_5 <{\overline s} g \sigma G s>+...\;\;. \label{4.7}
\eea
where the domain $D$ should now also satisfy the kinematical 
constraints specified below. 
After a double Borel transform in the variables $-q_1^2$
and $-q_2^2$, we obtain:
\bea
A\; F(q^2) m_{\phi} f_{\phi} &=&
e^{{m_{\phi}^2 \over M_1^2}} e^{{m_{\eta}^2 \over M_2^2}}
 \Bigg\{ \int d s_1 \int d s_2  
e^{-{s_1 \over M_1^2}} e^{-{s_2 \over M_2^2}}
 { 3 m_s \over \pi^2 \sqrt{\lambda(s_1,s_2,q^2)} }
\nonumber \\
+ e^{-{m_s^2\over M_1^2}}e^{-{m_s^2 \over M_2^2}}
 \Bigg[ <{\overline s} s> && \left(2 -{m_s^2 \over M_1^2}-{m_s^2 \over M_2^2}
+{m_s^4 \over M_1^4}+{m_s^4 \over M_2^4}+{m_s^2(2m_s^2-q^2)\over M_1^2
M_2^2} \right) \nonumber \\
+ <{\overline s } g \sigma G s>&& \left( {1 \over 6 M_1^2} +{2 \over 3 M_2^2}
-
{ m_s^2 \over 2 M_1^4}-{m_s^2 \over 2 M_2^4} +{(2 q^2 -3 m_s^2)
\over 3 M_1^2 M_2^2} \right) \Bigg] \Bigg\} \label{srb}
\eea
The integration domain $D$ over the variables $s_1,s_2$ depends on
the value of $q^2$ and is given by $D=D_1 \cup D_2$ where:
\begin{itemize}

\item $(-q^2)> s_{02}-4 m_s^2$

\hskip 3 cm
$D_1$:  $(s_2)_- \le s_2 \le s_{02}$  \hskip 1 cm $4 m_s^2 \le
s_1 \le
s_{01}$
\par

\item $(-q^2)< s_{02}-4 m_s^2$
\bea
D_2:  (s_2)_- \le s_2 \le (s_2)_+ \;\;\;\;\;&&4 m_s^2 \le s_1 \le
(s_1)_- 
\nonumber \\
(s_2)_- \le s_2 \le s_{02} \;\;\;\;\;&&(s_1)_- \le s_1 \le
s_{01}
\eea
\end{itemize}

\noindent with:
\be
(s_2)_\pm={2 m_s^2 q^2+(2 m_s^2-q^2) s_1 \pm \sqrt{s_1 q^2 (q^2-4
m_s^2)(s_1-4 m_s^2)} \over 2 m_s^2}
\ee

\be
(s_1)_\pm={2 m_s^2 q^2+(2 m_s^2-q^2) s_{02} \pm \sqrt{s_{02} q^2 (q^2-4 
m_s^2)(s_{02}-4 m_s^2)} \over 2 m_s^2} \;.
\ee

\EPSFIGURE{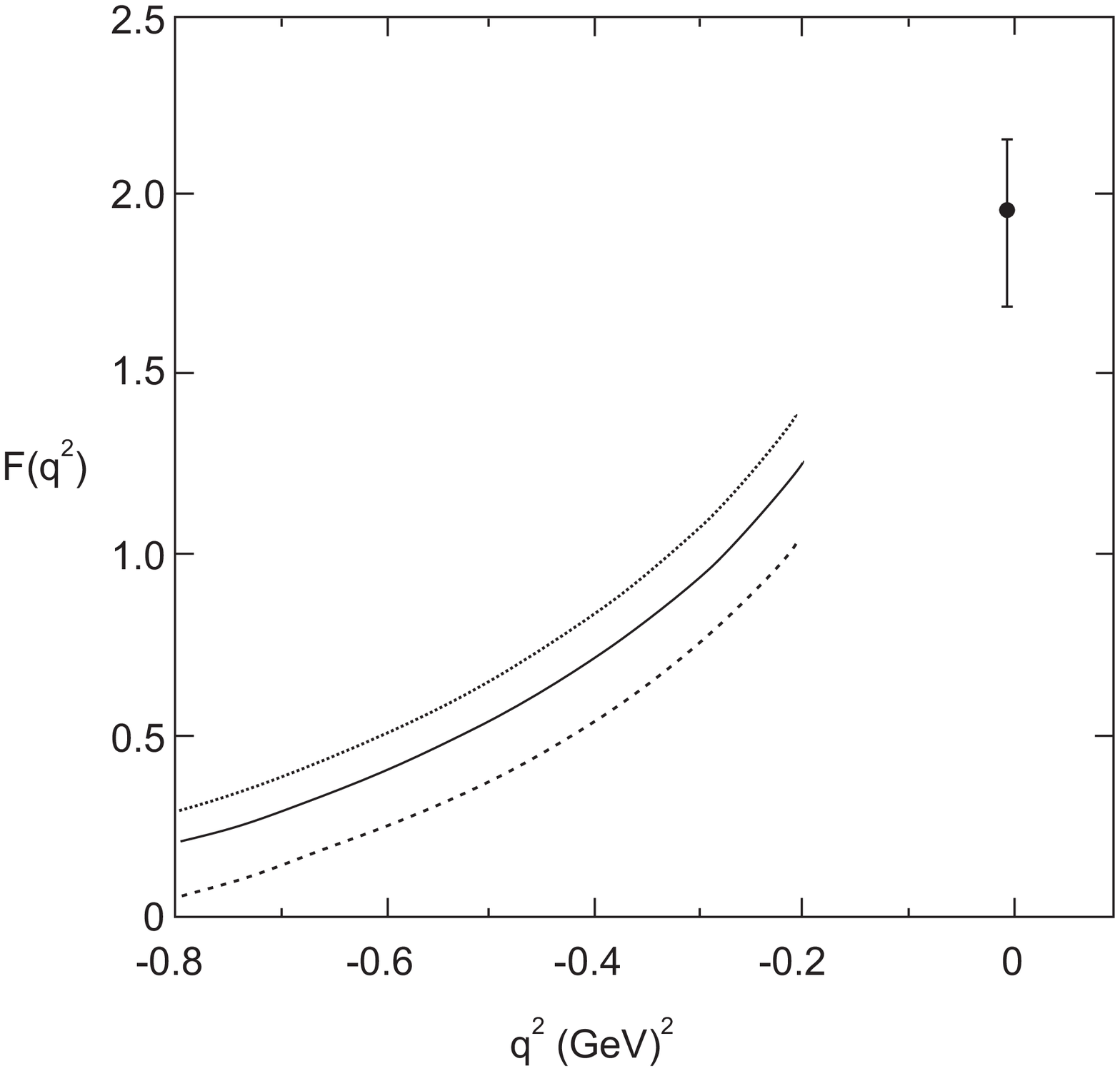,width=7.2cm}
{\label{fig:fq2} Form factor $F(q^2)$ obtained varying the input
parameters in the sum rule (\ref{srb}). The isolated point on the right is
the result of an extrapolation. The extrapolation of the solid curve gives
the central point on the right, corresponding to the result (\ref{g}).}

Since we  consider the form-factor $F(q^2)$ for arbitrary negative values of
$q^2$, we could perform a double Borel transform in the two
variables $Q_1^2=-q_1^2$ and $Q_2^2=-q_2^2$, which allows us to remove  
single poles in the $s_1$ and $s_2$ channels
(``parasitic'' terms) from the  sum-rule. Our procedure  
 is therefore to compute the form-factor $F(q^2)$ and
then to extrapolate the result to $q^2=0$. Strictly speaking, since we
only know the magnitude of $A$ in (\ref{4.7}), it is the modulus of
$F(q^2)$ that is determined.
In the numerical analysis we use:
$ m_\phi=1.02$ GeV,
$f_\phi=0.234$ GeV (obtained from the experimental datum on the decay to
$e^+
e^-$ \cite{pdg}).
We compute the result for  two values of
the $\phi$ threshold: $s_{01}=1.8,1.9$ GeV$^2$.
$s_{02}$ coincides with the $\eta$ threshold chosen as we did for the two
point function in section~3.

The outcome of the sum rule is depicted in figure~\ref{fig:fq2}. Varying
all
the parameters entering in the sum rule we obtain a region delimited by
the dashed and the dotted curves in this figure: they correspond to the
set of parameters giving the highest and the lowest curve analytically
defined by (\ref{srb}). The resulting form factor shows a behaviour in
$-q^2$ in the region $(-0.8,-0.2)$ GeV$^2$ which
can be fitted by a parabolic function. The extrapolation to $q^2=0$ gives:
\be
|g|={F(0) \over 3}=(0.66 \pm 0.06)\,{\rm GeV}^{-1} \; . \label{g}
\ee 
\noindent
The central value in (\ref{g}) corresponds to the extrapolation of the
solid line in figure~\ref{fig:fq2}. The uncertainty range in $|g|$, as
obtained by the extrapolation procedure, is displayed in the same figure. 

We can now use this result to compute the $\phi \to \eta \gamma$ decay width:
\be
\Gamma (\phi  \to \eta \gamma)={\alpha g^2 \over 24}
\Big({m_\phi^2-m_\eta^2 \over m_\phi } \Big)^3 \; ,
\ee
and, using $\Gamma(\phi)=4.43$ MeV \cite{pdg}, we obtain:
\be
{\cal B}(\phi \to \eta \gamma)  =(1.15 \pm 0.2)\% \; ,
\label{breta}
\ee
\noindent which compares favourably with 
the experimental outcome:
${\cal B}(\phi \to \eta \gamma) = (1.18 \pm 0.03 \pm 0.06) \%$
\cite{novos}.

We can extend the previous analysis to the channel with the $\eta^\prime$
in the final state. The derivation of the sum-rule is straightforward:

\bea
&& A^\prime \; F^\prime(q^2) m_{\phi} f_{\phi} e^{-{m_{\eta^\prime}^2
\over M_2^2}}+
A\; F(q^2) m_{\phi} f_{\phi} e^{-{m_{\eta}^2 \over M_2^2}}= \nonumber \\
&=&
e^{{m_{\phi}^2 \over M_1^2}} 
 \Bigg\{ \int d s_1 \int d s_2  
e^{-{s_1 \over M_1^2}} e^{-{s_2 \over M_2^2}}
 { 3 m_s \over \pi^2 \sqrt{\lambda(s_1,s_2,q^2)} }
\nonumber \\
&+& e^{-{m_s^2\over M_1^2}}e^{-{m_s^2 \over M_2^2}}
 \Bigg[ <{\overline s} s> \left(2 -{m_s^2 \over M_1^2}-{m_s^2 \over M_2^2}
+{m_s^4 \over M_1^4}+{m_s^4 \over M_2^4}+{m_s^2(2m_s^2-q^2)\over M_1^2
M_2^2} \right) \nonumber \\
&+& <{\overline s } g \sigma G s> \left( {1 \over 6 M_1^2} +{2 \over 3 M_2^2} -
{ m_s^2 \over 2 M_1^4}-{m_s^2 \over 2 M_2^4} +{(2 q^2 -3 m_s^2)
\over 3 M_1^2 M_2^2} \right) \Bigg] \Bigg\}
\eea

\noindent The integration region is the same as before 
with the  substitution:
$s_{02} \to s_{02}^\prime=s_0^\prime$ 
(as in the two point sum-rule of section~3).
We then obtain:
\be
|g^\prime|={F^\prime(0) \over 3}=(1.0 \pm 0.2)\; {\rm GeV}^{-1}
\label{gp}
\ee
\noindent
 which yields
\be
{\cal B}(\phi \to \eta^\prime \gamma) = (1.18 \pm 0.4) ~10^{-4}
\label{bretap} \; .
\ee
\noindent
The experimental datum is: 
${\cal B}(\phi \to \eta^\prime \gamma) = (0.82^{+0.21}_{-0.19} \pm
0.11)~10^{-4}$ \cite{novos}, completely compatible with our result
(\ref{bretap}).

These results have been derived without
including QCD radiative corrections. In principle, each term in the OPE
could be computed as an expansion in powers  of $\alpha_s$, supplementing
the
non-perturbative expansion with short-distance corrections. This would
display the correct scale and scheme dependence for the hadronic
quantities, such as  the coupling of the $\eta$ and $\eta^\prime$
to the currents considered in our analysis. 
The calculation of QCD corrections is a difficult task well
beyond the scope of the present paper. However, we would like to  comment
on the possible role of such terms. 

As far as the two point sum rule is concerned, ${\cal O}(\alpha_s)$
contributions have been computed \cite{chet}.
At a
typical scale $\mu=1$ GeV, these corrections are sizeable,
indicating that still higher orders may also be important. 
On the other hand, the main goal of the present analysis is the
computation of $\phi$ radiative decays. These results are obtained from
the
ratio of three point to two point sum rules. The most reliable procedure
in this case is to compute consistently the three point and two point
correlators at the same order in $\alpha_s$.
The uncertainty due to the neglect of higher order corrections should be
reduced due to a cancellation in the ratio.

This expectation is fulfilled, for example, in the calculation using QCD
sum
rules of the
Isgur-Wise function \cite{neubert}, describing in the heavy quark limit
the $B\to D^{(*)}$ semileptonic
transitions. In this case, though the ${\cal O}(\alpha_s)$ corrections are
large for the two point sum rules \cite{bagan},   explicit
calculation of the three point
correlator shows the expected cancellation,
i.e. the modest role of radiative corrections in the outcome. In this
particular case, the result is expected, at least at the zero recoil
point,  
by symmetry requirements. However, this cancellation
works in more general situations, such as the one presented in
\cite{noitau}, where the universal form factor describing the $B$
transitions to orbitally excited charmed mesons was computed at order
$\alpha_s$ by the same method. Again, although the two point correlator
received  important corrections, the ratio ot three to two point
functions is quite stable.

In the light of this discussion, our results for the
decay constants, eqs. (\ref{a}),(\ref{ap}), should be considered to be
estimates, the uncertainties in which
do not take into account possibly sizeable radiative corrections. On the
other hand, the outcome for radiative $\phi$ decays,
eqs. (\ref{breta}), (\ref{bretap}),
should be viewed as  much more accurate.

Indeed, the predictions for the branching ratios in (\ref{breta}) and
(\ref{bretap}) are the major
 results of this paper. They are quite independent 
of any mixing scheme for the $\eta$ and $\eta^{\prime}$.
Nevertheless, adopting the mixing scheme in the flavour basis 
described in section~2, it
is possible to derive the relation:
\be
R={ {\cal B}(\phi \to \eta \gamma)  \over  {\cal B}(\phi \to \eta^\prime 
\gamma)} =\Bigg({m_\phi^2-m_\eta^2 \over m_\phi^2-m_{\eta^\prime}^2}
\Bigg)^3 \tan^2 \phi_s 
\ee
from which we get $\phi_s=(34 \pm^8_6)^o$.
The experimental ratio would give: $\phi_s=(39.0 \pm^{7.5}_{5.5})^o$.

As  mentioned in the introduction,  the results obtained can, in
principle,
 provide us with  information about the magnitude of the WZW term, 
which represents an OZI-rule violating contribution to the effective
lagrangian. 
The strength of this term is
parametrized by a constant $\Lambda_3$ and is determined by    the values
of the  couplings $g$ and $g^\prime$. For example, in
\cite{feldrev} it is found:
\be
g={3 m_\phi \over 2 \pi^2 f_\phi} \left( {1 \over 6} {\cos \phi_s \sin
\phi_V \over f_q}+{1 \over 3} {\sin \phi_q \over f_s} +{\Lambda_3 \over 3
\sqrt{3}} {\sin \theta_8 \over f_0} \right) \;\;,
\label{phil3}
\ee
\noindent where $\phi_V$  is the mixing angle in the $\phi-\omega$
system.
We  assume this formula to estimate the size of
$\Lambda_3$. Unfortunately,  the combined effect of the
uncertainties
affecting the parameters entering   (\ref{phil3}) allows us no more
definite conclusion than 
 $\Lambda_3 \simeq 2 \pm 4$. Precise measurements of $\phi$ radiative
decays to $\eta$ and $\eta^\prime$ at KLOE \cite{book} will do better.

Let us now compare our results with previous determinations. In
ref. \cite{ball} the chiral anomaly prediction at $q^2=0$ and the vector
meson dominance are exploited to  derive the couplings $g=g_{\phi \eta
\gamma}$ and $g^\prime=g_{\phi \eta^\prime \gamma}$. A single angle mixing
framework in the octet-singlet basis is assumed and the corresponding
coupling constants $f_0$, $f_8$ are derived from the experimental data on
the decays $\eta \to \gamma \gamma$ and $\eta^\prime \to \gamma \gamma$ and
used as an input to derive $g$ and $g^\prime$ as a function of the mixing
angle. 
In  \cite{feld} this approach is extended to the quark-flavour mixing
scheme
with the result: $g=0.78$ GeV$^{-1}$ and $g^\prime=0.95$ GeV$^{-1}$.
On the other hand, an energy-dependent mixing scheme is adopted in
\cite{escribano}, with the
result: $g=(0.73 \pm 0.06)$ GeV$^{-1}$ 
and $g^\prime=(0.83 \pm 0.06)$ GeV$^{-1}$. 
Alternatively, Ref. \cite{benayoun} exploits the hidden local symmetry 
approach, together
with the inclusion of various   $SU(3)$ symmetry breaking \cite{bky,bgp} to
obtain $g=g^\prime=0.70$ GeV$^{-1}$. As can be observed, the
various approaches seem to agree quite well for the decay with the 
$\eta$ in the final state, while the results have a larger spread in the
case of the $\eta^\prime$. For a more comprehensive survey of results
we refer to \cite{feldrev}.

\section{  $\eta$ and   $\eta^\prime$ couplings to axial-vector currents using
two-point sum-rules }

Our study of $\phi$ radiative decays to $\eta$ and $\eta^{\prime}$ 
require no knowledge of $\eta-\eta'$ mixing, only the coupling to strange quarks is needed. However, using similar techniques, we can 
investigate their decay constants
in both the strange and non-strange sectors 
and so deduce the mixing pattern within errors. 
This is the purpose of this section. We begin by considering
the correlator
\be
\Pi_{\mu \nu}(q^2)=i ~\int d^4 x~ e^{iq \cdot x} <0|T[ J_{5
\mu}^s(x) J_{5 \nu}^s(0)]|0>
\ee
Following the procedure already outlined above, we obtain the
sum-rule:

\be
(f_\eta^s)^2=e^{m_\eta^2 \over M^2}  \Bigg[ {1 \over \pi} \int_{4
m_s^2}^{s_0}
ds ~e^{-{s \over M^2}} \rho^{pert}(s) +{2 m_s \over M^2} <{\overline
s}s> e^{-{m_s^2 \over M^2}}
\Bigg]
\ee

\noindent where:
\be
\rho^{pert}(s)={1 \over 4 \pi} \sqrt{1 -{4 m_s^2 \over s}}\quad {2 m_s^2+s
\over s}
\ee
(in this case the $d=5$ contribution vanishes).

\EPSFIGURE{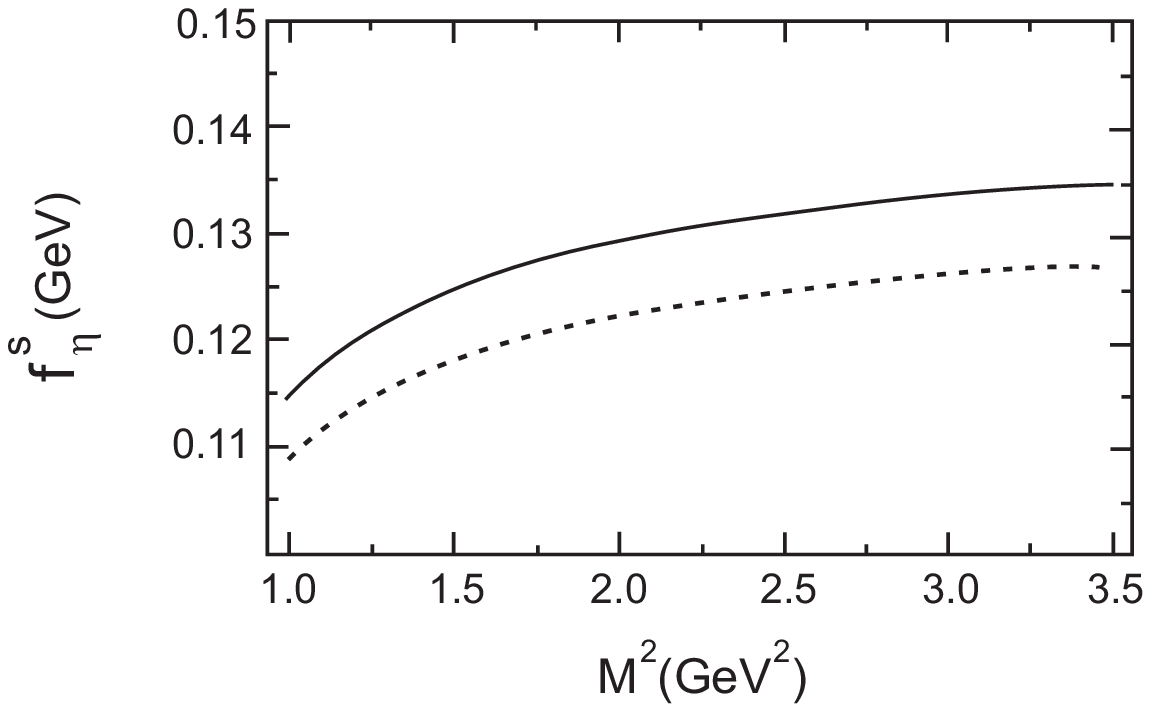,width=7.2cm}
{\label{fig:fetas}
Constant  $f_\eta^s$ as a function of the Borel parameter $M$ for 
$m_s=0.133$ GeV. The solid  curve
corresponds to $s_0$=(0.95 GeV)$^2$, the dashed 
one to $s_0$=(0.9 GeV)$^2$.}

The allowed range for the Borel parameter, obtained according to the above
criteria, is:
 $0.35$ GeV$^2 \le M^2 \le 3.5$ GeV$^2$, and the  
further stability window is found  in $[2,3.5]$ GeV$^2$.
The result is depicted in figure~\ref{fig:fetas}, for the value
$m_s=0.133$ GeV. Taking into account the uncertainty in $m_s$ too we get:
 \be
f_\eta^s=(0.13 \pm 0.01)\; {\rm GeV} \;.
\label{fetas}
\ee 

In order to determine $f_{\eta^\prime}^s$ one has to repeat the
previous calculation more or less exactly, 
raising the threshold above the $\eta^\prime$ mass and
considering the pole contribution of the $\eta$ on the hadronic side of
the sum-rule.
The result is:
\be
(f_{\eta^\prime}^s)^2 e^{-{m_{\eta^\prime}^2 \over M^2}}+  
(f_\eta^s)^2 e^{-{m_\eta^2 \over M^2}}  =  {1 \over \pi} \int_{4
m_s^2}^{s_0^\prime}
ds\, e^{-{s \over M^2}} \rho^{pert}(s) +{2 m_s \over M^2} <{\overline
s}s> e^{-{m_s^2 \over M^2}}
\ee
where $\rho^{pert}(s)$ is the same as before.
In the numerical analysis,  $s_0^\prime$ is varied in the range
$(1.2-1.25$~GeV)$^2$. The
selected stability window for $M^2$ is $[2,5]$ GeV$^2$.
We obtain:
\be
f_{\eta^\prime}^s=(0.12 \pm 0.02)\; {\rm GeV} \; .
\label{fetaps}
\ee

If we now  use the current $J_{5 \mu}^q$ in the correlator,  instead of
$J_{5 \mu}^s$, and
 we set the up and down quark masses to zero, we can obtain $f_{\eta}^q$.
The stability window is found for $M^2$ in the range $[2,4]$ GeV$^2$ with the result
\be
f_\eta^q=(0.144 \pm 0.004)\;{\rm GeV} \;. \label{fetaq}
\ee 
 Raising the threshold, we can also evaluate
$f_{\eta^\prime}^q$, with the result:
\be
f_{\eta^\prime}^q=(0.125 \pm 0.015)\;{\rm GeV} \;. \label{fetapq}
\ee
\noindent 
As  with the results obtained in section~3 from two-point sum
rules,  these also represent estimates,  derived without
radiative QCD corrections.

We can now use the set of results (\ref{fetas}), (\ref{fetaps}),
(\ref{fetaq}), (\ref{fetapq}) to
estimate all the mixing parameters appearing in (\ref{constmix}).
From the relation ${f_{\eta}^s/ f_{\eta^\prime}^s}=\tan \phi_s$, 
one has $\phi_s=(46.6^o\pm 7^o)$.
Since  $f_\eta^s=f_s \sin \phi_s$ \footnote{This relation is to be
considered in terms of absolute values, since the sum-rule gives access
only to $(f_\eta^s)^2$, and therefore does not allow the
sign of $f_\eta^s$ to be determined.}, a prediction for $f_s$ follows:  
$f_s=(0.178 \pm 0.004)$~GeV, the central value corresponding to $f_s=1.345
f_\pi$, with $f_\pi=0.132$~GeV.

Using the relation:  
${f_{\eta^\prime}^q/f_{\eta}^q}=\tan \phi_q$, we find $\phi_q=(41^0 \pm
4^o)$.
We can now derive a prediction for $f_q$,
using $f_{\eta}^q=f_q~\cos\phi_q$. We obtain $f_q=(0.19 \pm 0.015)$~GeV,
the
central value of which corresponds to $f_q \simeq 1.44 f_\pi$. 
Let us observe that our results correspond to  
$|\phi_s -\phi_q|/(\phi_s +\phi_q) \simeq 0.065$, 
which confirms the relation put
forward in \cite{feld} that this ratio should be much less than 1. 

If we now turn to the scheme with two mixing angles in the octet-singlet basis,
we could exploit the relations (\ref{twoangles})-(\ref{f08}) to obtain:
\bea
\theta_8 \simeq -8.4^o \hskip 1 cm && \theta_0 \simeq -13.8^o \nonumber \\
f_8 \simeq 1.44~f_\pi \hskip 1 cm && f_0 \simeq 1.35~f_\pi \;.
\label{octris}
\eea

Previous determinations of the parameters calculated above range over
large intervals, expecially for those corresponding to  
(\ref{octris}),
i.e. in the octet-singlet mixing scheme. For a comprehensive collection of
previous
results we again refer  to \cite{feldrev}. We only observe that our results
for
$\phi_q$, $\phi_s$ are in pretty good agreement with those in
refs. \cite{kaiser,benayoun,schechter}. Our outcome for $f_s$ also agrees
quite well with most of
previous results \cite{feldrev}, while the result for $f_q$ seems
somewhat larger than previous determinations.

We can now exploit the results obtained in this section, i.e. the values
in  (\ref{fetas}), (\ref{fetaps}), together with the predictions
(\ref{a}) and (\ref{ap}), to derive the following matrix elements from
(\ref{glue}):
\bea
\langle 0 \left|{\alpha_s \over 4 \pi}G \tilde G\right
|\eta\rangle &=&(0.008 \pm 0.004)\,{\rm GeV}^3 \quad ,\nonumber \\
\langle 0 \left|{\alpha_s \over 4 \pi}G \tilde G
\right|\eta^{\prime} \rangle &=&(0.072 \pm 0.025)\,{\rm GeV}^3
\label{gg} \;.
\eea
\noindent
Both values in (\ref{gg}) are close to the naive quark model 
calculation of Novikov et al. \cite{novikov}, 
particularly that for the $\eta^{\prime}$, which is
$f_{\pi}\ m_{\eta^{\prime}}^2/\sqrt{3}= 0.070$~GeV$^3$. 
Our result for the $\eta$ is somewhat smaller than the one found in
\cite{novikov}:
$f_{\pi}\ m_{\eta}^2/\sqrt{6}= 0.016$~GeV$^3$.
However, their simple quark model result does not take into account
$SU(3)_F$ breaking corrections. In contrast, ours does. 
Thus, the  matrix elements of  (\ref{gg}) are important for the 
investigation of the structure
of the $\eta$ and $\eta^\prime$ 
and their possible 
glue content \cite{novikov,frere}.

\section{Conclusions}

We have analysed radiative $\phi \to \eta \gamma$, $\phi \to \eta^\prime
\gamma$ decays using QCD sum-rules. This analysis required a preliminary
calculation of the couplings of the pseudoscalar current to the $\eta$ and
$\eta^\prime$. The sum-rules are derived without any assumption 
about $\eta-\eta^\prime$ mixing. 
Though we use only lowest order in QCD perturbation theory, potentially
large higher order corrections are expected to cancel between the three
and two point correlators, to give results that are  in
good agreement with the available experimental data on the $\eta$ channel,
and   are compatible with the Novosibirsk datum for the
$\eta^\prime$ case.
Since the 
uncertainty in the latter is large, the last word is still 
left to the experimental
improvement  at DA$\Phi$NE, for instance. 

We have also discussed the issue of $\eta-\eta^\prime$
mixing, giving
predictions for the parameters describing such mixing in a quark-flavour
basis  scheme. We observe that the two angles required in such a scheme are
quite close to each other. The existing
spread of results gives us
confidence that new experimental information will shed light on this
sector of low-energy physics too.
\vspace{2cm}

\noindent{\bf Acknowledgments}

We are most grateful for support from the EU-TMR Programme, 
Contract No.
CT98-0169, EuroDA$\Phi$NE.

\end{document}